\documentclass{article}

\newcommand{\etal}{{\it et al.}}
\newcommand{\be}{\begin{equation}}
\newcommand{\ee}{\end{equation}}
\newcommand{\bea}{\begin{eqnarray}}
\newcommand{\eea}{\end{eqnarray}}

\renewcommand\[{\left[}

\begin{document}\date{}
\title{A bound on galactic mass loss rates obtaind from globular cluster dynamics}
\author{Florian Dubath$^\dag$  and Anna Rissone$^\dag$\\
{\scriptsize$^\dag$ D\'epartement de Physique Th\'eorique, Universit\'e de Gen\`eve,}\\{\scriptsize 24 quai Ernest-Ansermet, 1211 Gen\`eve 4, Switzerland}\\
{\scriptsize florian.dubath@physics.unige.ch}}

\maketitle
\begin{abstract}
Using tidal disruption of globular clusters by the galactic center, we put limits on the total mass ever enclosed into orbits of observed globular clusters. Under the assumption that the rate of mass loss from the Galaxy is steady, we then deduce a bound on this rate. In particular this bound can be used to constrain the galactic gravitational wave luminosity.
\end{abstract}

\section{Introduction}
Upper bounds on the mass loss of the Galaxy have been discussed in the past, especially in connection with the maximum luminosity of the Galaxy in gravitational waves \cite{sciama,poveda}. Recently, in Ref. \cite{CDM}, the issue has been revisited, partly motivated by the 2001 data of the NAUTILUS/EXPLORER collaboration \cite{astone}. As discussed in \cite{CDM}, limits on the mass loss are given by its effect on galactic dynamics. If the Galaxy loses mass, it becomes less bound and expands. One can constrain this expansion from various observations; in particular:\\
  (i) The radial velocity of stars in the sun's neighborhood. In particular a component linear with the distance and independent of the direction (i.e. a K-term in the galactic dynamics literature).\\
  (ii) The relative velocity of this neighborhood with respect to the galactic center.\\
  (iii) The existence of old wide binaries near the galactic center.\\
  (iv) The existence of globular clusters near the galactic center.\\
This last method, based on tidal disruption of globular clusters, was proposed by A. Poveda and C. Allen \cite{poveda} and applied to the globular cluster Omega Centauri. Actually some facts about its metalicity, mass and shape support the idea that $\omega$ cen is a rather unusual globular cluster and probably the remains of a tidally stripped dwarf galaxy \cite{Majewski,tsuchiya,Carraro,Fellhauer}. In this case, it would not satisfy the assumption of an adiabatically evolving orbit required in the analysis of Ref. \cite{poveda}. For this reason we repeat this analysis using other globular clusters.\\
We first discuss in Sec.\ref{EO} the evolution of elliptical orbit in the potential of a varying mass. Then we recall some facts about the tidal radius and the dynamical evolution of globular clusters. In Sec.\ref{LL}, combining this and the observations, we are able to derive an upper limit on the mass loss.

\section{Evolution of the globular clusters \label{EO}}
In this section we derive the tidal radius of globular clusters using Newtonian gravity and disregarding other effects.\\
If there is a continuous mass loss in the Galaxy, stellar orbits change with
time. However, for slow changes of the galactic mass, there exist some adiabatic invariants in the description of the orbit (see \cite{binney} Chap. 3).
In the case of an ellipse around a mass $M$, the eccentricity $e$ and the product of the mass with the major semi axis, $Ma$, are such invariants. This adiabatic invariance property can be then extended to more
general cases \cite{lynden-bell}. For an
axially symmetric field and orbits on the galactic plane the adiabatic invariance
of the eccentricity is still valid (with a more general definition for
the eccentricity for non elliptical orbits). In particular, if we
consider the osculatory ellipse at perigalacticon\footnote{For non elliptical orbits the osculatory ellipse at perigalacticon is the ellipse approximating the orbit near this point.}, we have $M = M_p$ (where $M_p$ denotes the mass
enclosed in a sphere of radius $R_p$), and the adiabatic invariance
means $e$ = constant and $M_p a $ =  constant, from which follows also
$M_p R_p$ = constant, which is the formula we will use in the following.
As mentioned above, this is correct if the motion lies in the
galactic plane; in general this is not true for globular cluster
orbits. However, if the orbit is of a box type there is on average
no exchange of energy between the vertical motion and the motion in
the plane \cite{ollongren}: the two motions can thus be separated and
the invariance theorem can be applied to the motion in the plane.
For this reason we consider in the following only globular clusters
which have box orbits.
A star cluster orbiting in a potential is subject to tidal forces. Due to
these forces the cluster could lose its most distant stars. We follow King's definition of the tidal radius \cite{king}. Let's
consider a spherical globular cluster of mass $M_c$ describing an
ellipse around the galactic center. The acceleration of the cluster at a distance $R$
with respect to the galactic center is given by:
\begin{equation}
\frac{d^2R}{d t^2}=R\omega^2-\frac{d V}{d R}\ ,
\end{equation}
where $\omega$ is the angular velocity and $V$ the galactic potential.
For a generic star in the cluster, the acceleration is:
\begin{equation}
\frac{d^2R_*}{d t^2}=R_*\omega^2-\left(\frac{d V}{d
R}\right){\!\Bigg\arrowvert}_{R_*}-G\frac{M_c(R_*-R)}{|R-R_*|^3} \ ,
\end{equation} where $R_*$ is the position of the star measured from the galactic center.\\
Using a Taylor expansion we can compute the relative acceleration to first order in $(R_*-R)$:
\begin{equation}
\frac{d^2R_*}{d t^2}-\frac{d^2R}{d
t^2}\cong\left(\omega^2-\frac{d^2V}{d R^2}-\frac{GM_c}{|R_*-R|^3}
\right)(R_*-R) \ .
\end{equation}
The tidal radius is defined as the distance from the cluster center
for which the relative acceleration vanishes (it corresponds to the
maximum distance that a star can reach without escaping from the cluster):
\begin{equation}
r_t=\left(\frac{GM_c}{\omega^2-\frac{d^2V}{d R^2}} \right)^{1/3}\ .
\end{equation}
Indicating with $M_{GC}$ the galactic mass inside a radius $R$ and assuming that it is spherically distributed we have:
\begin{equation}
\frac{d^2V}{d R^2}=-2G\frac{M_{GC}}{R^3}\ .
\end{equation}
Using the geometry of the ellipse,
$\omega^2=GM_{GC}R^{-4}a(1-e^2)$, we obtain for the tidal radius at
perigalacticon\footnote{The tidal radius at perigalacticon can
be considered the effective tidal radius of the cluster: for a
discussion of this issue see \cite{king, hoerner}}
($R=R_p\equiv a(1-e)$, $M_{GC}=M_p$):
\begin{equation}
r_t=R_p\left( \frac{M_c}{M_{p}(3+e)}\right)^{1/3}\ .
\end{equation}

In the following we make use of a well known fact about globular clusters evolution:
in the first phase of the evolution, until core collapse, the
central part of a 
globular cluster undergoes a dynamical contraction (see for example
\cite{spitzer1}-\cite{meylan}) while the
outer part expands. The globular clusters that we are considering in
our analysis are all in this evolutionary phase.
It remains difficult to determine the fraction of the mass that makes the
contraction. 
Poveda and Allen \cite{poveda} estimated it to be $75\%$ making use of
results found by Spitzer and al. \cite{spitzer1, spitzer2} using numerical simulations. According to more recent estimations
\cite{elson, meylan2} we consider the smaller
fraction $1/2$. Actually for our purpose the only thing we need to
consider is a fraction of the cluster mass which undergoes a contraction. It is
clear that if we consider the maximal fraction with this property we
obtain the tightest bound on the mass loss; but in any case even with a smaller
fraction we obtain a limit, even if is less strict. This explain why we obtain a different bound than Poveda and Allen for $\omega$cen.

\section{Limit on the total mass loss\label{LL}}
The tidal radius of globular clusters gives informations on the mass of the galactic center enclosed in a sphere of radius $R_p$. By mass we mean total gravitational mass and therefore the analysis makes no difference between "normal" and "dark" matter. Furthermore the limit on mass loss gives no contraint on the nature of the mass lost or on the loss mechanism.\\
Following the idea of Poveda and Allen \cite{poveda}, we can obtain a limit
on the total mass loss, assuming at first that the mass loss is localized at the galactic center.
As described above, the radius $r_x$, containing a
fraction $x$ of the cluster mass, decreases with time for
small enough $x$. We have the inequality:
\begin{equation}
r_t^i=\alpha r_x^i\geq\alpha r_x^n\ ,
\end{equation}
where $\alpha$ is the quotient between the tidal radius $r_t$
and $r_x$, the superscripts $i$ and $n$ stand for "initially" and "now"
respectively.\\
Using the formula for the tidal radius and the adiabatic invariance of
$M_p R_p$ and $e$ we have:
\begin{equation}
\frac{R_p^n M_p^n}{M_p^i}\left
( \frac{M_c}{(3+e)M_p^i}\right)^{1/3}\geq\alpha r_x^n \ .
\end{equation}
From this equation we can derive an upper limit on $M_p^i$.
In order to do this, we should use a model for the mass distribution to determine
$\alpha$, but it is clear that it is in any case smaller than for the case of a
uniform density distribution, for which $\alpha = 1/x^{1/3}$.
Thus we have:
\begin{equation}
M_p^i\leq\left( \frac{R_p^n M_p^n}{\frac{1}{x^{1/3}}
r_x^n}\right)^{3/4}\left(\frac{M_c}{3+e} \right)^{1/4} \ .\label{MPI}
\end{equation}

Since all the galaxy has expanded then $M_p^n$, the mass enclosed in a
sphere of radius $R_p^n$, was originally enclosed in a sphere
of radius $R_p^i$ and therefore the mass loss is given by:
\begin{equation}
\Delta M =M_p^i-M_p^n\ .
\end{equation}

\subsection{Results}
The most stringent limits correspond to clusters with small
peri galacticon radius. On the other hand clusters with $R_p\leq 1[kpc]$
have chaotic orbits and do not satisfy the adiabatic invariance conditions.
We restrict our choice to clusters known to have box orbits, see \cite{Allen}, but
with $R_p$ small enough; in all the cases that we consider, $R_p$ is
approximatively the size of the bulge, so we can take $M_p$ as the
mass of the bulge.

Orbit type determination are obtained taking a fixed galactic potential (this is also the case in \cite{Allen}). It would be interesting to study how the result changes if one use a time-varying potential wich takes into acount the variation of the mass of the Galaxy. However, we do not expect substential changes in the final result because of the structure of equation (\ref{MPI}). In fact a miss-determination of $R_p^n$ and $e$ of said 20\% do not change the order of magnitude of our result.

The parameters that we need for our selected clusters are:
\begin{center}
\begin{tabular}{|c|c|c|c|c|c|}
\hline Name & NGC & $M_c\ [M_{\odot}]$ (\cite{meziane}) & $R_p^n\
[kpc]$ (\cite{dinescu}) & e (\cite{dinescu}) & $r_{0.5}\ [pc]$
(\cite{trager})  \\ \hline \hline
$\omega$ cen &5139&$2.40\times 10^{6}$&1.2&0.67&11.7\\ \hline
M107&6171&$6.31\times 10^{4}$&2.3&0.21&7.54\\ \hline
&6218&$1.17\times 10^{5}$&2.6&0.34&7.96\\ \hline
M92 &6341&$2.19\times 10^5$&1.4&0.76&5.75\\ \hline
M28 &6626&$2.29\times 10^{5}$&2.1&0.19&7.17\\ \hline
\end{tabular}
\end{center}
With $M_B\cong1.6\times 10^{10}[M_{\odot}]$ the mass of the galactic
bulge, we obtain
\begin{center}
\begin{tabular}{|c|c|c|c|c|c|}\hline
&$\omega$ cen&M107&NGC6218&M92&M28\\ \hline
&&&&&\\[-3mm]
$\Delta M\ [M_{\odot}]\leq$&$1.9\times 10^{10}$&$1.7\times 10^{10}$&$2.4\times 10^{10}$&$2.0\times 10^{10}$&$2.7\times 10^{10}$\\ \hline
\end{tabular}
\end{center}
If the mass is only lost from the center of the galaxy, then this is the
total mass loss. For a steady loss during the whole lifetime of the
galaxy (and taking the lifetime of the galaxy to be $\Delta t = 1.2\times
10^{10} [yr]$) the maximum rate would be
\begin{center}
\begin{tabular}{|c|c|c|c|c|c|}\hline
~~~~~~~~~&~~~~$\omega$ cen~~~&~~~M107~~~&NGC6218&~~~~M92~~~~&~~~~M28~~~~\\ \hline
&&&&&\\[-3mm]
$ \frac{\Delta M}{\Delta t_{~\!}}[M_{\odot}/yr]\leq $&1.6&1.4&2.0&1.7&2.3\\\hline
\end{tabular}
\end{center}
From the above analysis we see that only the variation of the mass inside the globular cluster orbits is taken into account, see eq.(\ref{MPI}). We have assumed that the mass loss takes place only there. For example the mass loss could be proportional to the luminous mass distribution. In this case, the total mass loss is given by
\be
\Delta M_{tot} = \frac{M_{tot}}{M_B} \Delta M
\ee
where $M_{tot}$ denotes the total luminous mass in the
galaxy, $\Delta M$ the mass loss in the bulge previously calculated,
and $\Delta M_{tot}$ the total mass loss in this case.
Taking $M_{tot} = 1.1\times 10^{11}[M_{\odot}]$  we obtain, assuming a steady loss during the whole lifetime of the
galaxy, the
limits on the rate of mass loss
\begin{center}
\begin{tabular}{|c|c|c|c|c|c|}\hline
~~~~~~~~~&~~~~$\omega$ cen~~~&~~~M107~~~&NGC6218&~~~~M92~~~~&~~~~M28~~~~\\ \hline
&&&&&\\[-3mm]
$ \frac{\Delta M_{tot}}{\Delta t_{~\!}}[M_{\odot}/yr]\leq $&11&9.7&14&11&16\\\hline
\end{tabular}
\end{center}

\section{Conclusion}
Globular clusters allow to fix a limit on (a steady) galactic mass loss of order of 1$[M_{\odot}/  yr]$ for the galactic center (central $kpc$) or, if we assume that the mass loss is correlated with luminous mass, of order of 10$[M_{\odot}/  yr]$ for the whole galaxy. These limits are independent of the kind of mass loss and then could be used to constrain any mechanism producing a net outflow of gravitational mass.\\
In particular the first limit gives a bound on the maximal gravitational wave luminosity from the galactic center. In \cite{astone} has been reported the analysis of the data of the 2001 run of the EXPLORER and NAUTILUS gravitational wave detector. More statistics is needed to draw any conclusion (and further runs are presently under way), but in that data there was a hint of a possible coincidences eccess when the bars are oriented favorably with respect to the galactic plane. If interpreted in term of gravitational wave emission from galactic sources, this would correspond to a mass loss of the Galaxy to gravitational waves of the order of $(2-6)[M_\odot/yr]$, see \cite{CDM}. Combined with other arguments (see \cite{CDM} for a detailed discution), our work shows that a model with sources of gravitational wave located in the galactic center, radiating steadliy for a time comparable with the age of the Galaxy, is only marginaly consistant with the observation. For this reason in \cite{CDM} have been considered other alternatives, like a single or just a few close sources emiting repeatesdly gravitational wave burst.

\section*{Acknowledgments}

We wish to thank Michele Maggiore, Stefano Foffa, Alice Gasparini, Riccardo Sturani and Samuel Leach for many useful discussions. This work was partially supported by the Fonds National Suisse.

\end{document}